\documentclass{elsart}

\usepackage{natbib}

 \usepackage{graphicx}

\usepackage{amssymb}

\begin{document}

\begin{frontmatter}



\title{Structure of $n$-clique networks embedded in a complex network}


\author[BIC,corr]{Kazuhiro Takemoto}
\ead{takemoto@kuicr.kyoto-u.ac.jp}
\author[BIO,BAP]{Chikoo Oosawa}
\author[BIC]{Tatsuya Akutsu}

\address[BIC]{Bioinformatics Center, Institute for Chemical Research, Kyoto University, Gokasho, Uji, Kyoto 611-0011, Japan}

\address[BIO]{Department of Bioscience and Bioinformatics, Kyushu Institute of Technology, Iizuka, Fukuoka 820-8502, Japan}

\address[BAP]{Bioalgorithm Project, Faculty of Computer Science and Systems Engineering, Kyushu Institute of Technology, Iizuka, Fukuoka 820-8502, Japan}

\corauth[corr]{Corresponding author.}

\begin{abstract}
We propose the $n$-clique network as a powerful tool for understanding global structures of combined highly-interconnected subgraphs, and provide theoretical predictions for statistical properties of the $n$-clique networks embedded in a complex network using the degree distribution and the clustering spectrum. Furthermore, using our theoretical predictions, we find that the statistical properties are invariant between $3$-clique networks and original networks for several observable real-world networks with the scale-free connectivity and the hierarchical modularity. The result implies that structural properties are identical between the $3$-clique networks and the original networks.
\end{abstract}


\end{frontmatter}

\section{Introduction}
Cliques are highly-interconnected subgraphs (complete graphs), and appear dominantly in networks which describe wide-ranging complex systems occurring from the level of cells to society. And, the cliques are actively investigated in recent years because of provisions of important insights to information processing, hierarchical modularity, and community structures. For instance, in gene regulatory networks, small cliques correspond to the feed-forward loop which is one of the network motifs \cite{Milo2002}. The motifs play an important role in gene regulation \cite{Alon2002}, and are regarded as building blocks of life. Furthermore, the cliques are a representation for clusters, communities, and groups \cite{Scotts2000, Watts2002} because there are edges among persons as nodes if there are friendships, partnerships, and {\it etc.} among the persons in social networks. Therefore, the cliques help to detect community structures \cite{Palla2005} in social networks.
Again, in protein-protein interaction networks, the cliques are powerful tools for understanding evolution of proteins and functional predictions of proteins having unknown function \cite{Palla2005} because proteins which have same functions tend to interact. 

Motivated by these breakthroughs, recent efforts have taken place to analytically evaluate the abundance of subgraphs, including cliques, based on statistical mechanics \cite{Bianconi2003,Itzkovitz2003}, providing excellent knowledge about the local interaction patterns \cite{Vazquez2004} and the time evolution of the abundance of subgraphs including cliques \cite{Vazquez2005}.
These previous works focus on the local information such as the subgraph and clique abundance, and the size of the giant components led by percolation via a class of subgraphs such as the subgraph percolation \cite{Vazquez2004}, the $L-$percolation \cite{Costa2004}, and the clique percolations \cite{Derenyi2005}. In recent years, however, it has been revealed that real-world networks are constructed by overlapping subgraphs including cliques \cite{Vazquez2004,Palla2005}; thus it is important to elucidate global structures in networks consisting of cliques. For example, dynamics of a high order emerge by the combined network motifs in gene regulatory networks \cite{Kashtan2004,Ishihara2005}.

In particular, the several power-law statistical properties have been empirically found in real-world complex networks. One of the properties is scale-free connectivity \cite{Barabasi1999} which is characterized by a power-law degree distribution $P(k)\sim k^{-\gamma}$ with $2<\gamma<3$ empirically found \cite{Reka2002}. The scale-free connectivity means that a few nodes (hubs) integrate a great number of nodes and most of the remaining nodes do not. Another of the properties is hierarchical modularity which is characterized by a power-law clustering spectrum $C(k)\sim k^{-\alpha}$ with $\alpha\approx 1$ empirically found, and this property suggests a hierarchical structure of the cliques \cite{Ravasz2003,Takemoto2005}. A clustering spectrum is defined as an average clustering coefficient of nodes with degree $k$, where the clustering coefficient means the density of edges among neighbors of a node. Since these properties reflect a global structure of a network, it is significant to clarify relationships between these properties and the global structures of the combined cliques. 

In this paper, we propose the $n$-clique network as a powerful tool for understanding global structures of combined highly-interconnected subgraphs. Furthermore, we provide the theoretical predictions for well-known statistical properties of $n$-clique networks embedded in a complex network using the degree distribution and the clustering spectrum, and evaluate our theoretical predictions with numerical simulations.
The theoretical predictions are established by applying the statistical method in \cite{Vazquez2005}. Moreover, we discuss relationships of statistical properties which are observed between several real-world networks and their $n$-clique networks.

\section{$n$-clique networks}
\begin{figure}[bp]
\begin{center}
	\includegraphics{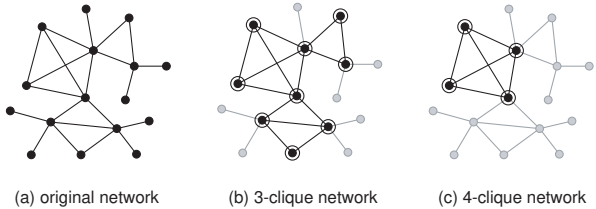}  
	\caption{Schematic diagram of $n$-clique networks embedded in the original network (a). The $n$-clique networks [(b) and (c)] are expressed as the circled black nodes with black edges.}
	\label{fig:c_net}
\end{center}
\end{figure}

$n$-clique networks are represented as sets of nodes and edges which are contained in $n$-node cliques, corresponding to $n$-node complete graphs, embedded in an original network.
Figure \ref{fig:c_net} shows a schematic diagram of $n$-clique networks.
The original network [Fig. \ref{fig:c_net} (a)] has two clique networks [Figs. \ref{fig:c_net} (b) and (c)], and the clique networks are expressed as the circled black nodes with black edges.
The gray nodes and edges are eliminated because the nodes and edges are affiliated with no cliques.
Following a procedure, $n$-clique networks are extracted from an original network.
In addition, original networks are equivalent to $2$-clique networks in the absence of isolated nodes corresponding to nodes which have no edges. In this paper, we assume that the original networks have no isolated nodes. We utilize the algorithm based on the network motif detection \cite{Milo2002} to find the cliques
Although finding clique abundance is computationally intractable (NP-hard), enumeration of $n$-cliques in a given network can be done in polynomial time if $n$ is a constant \cite{Skiena1997}.

\section{Degree distribution}
\begin{figure}[tbp]
\begin{center}
	\includegraphics{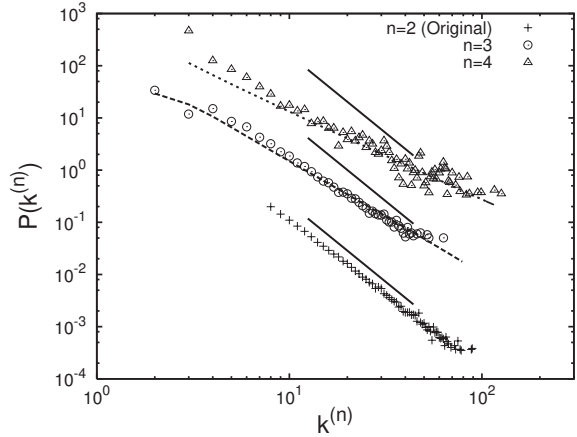}  
	\caption{Degree distributions of $n$-clique networks embedded in the BA network with $N=3,000$ and $\langle k \rangle=16$ (shifted for clarity). $\langle k \rangle$ means the average degree. The symbols correspond to the numerical results, and the dashed lines are theoretical predictions given by Eq. (\ref{eq:P_n(k)}). The solid lines show $P(k)\propto k^{-3}$.}
	\label{fig:deg_BA}
\end{center}
\end{figure}

We consider degree distributions from $n$-clique networks $P(k^{(n)})$. The degree distribution is defined as the existence probability of nodes with degree $k^{(n)}$ which is the number of edges at a node in a $n$-clique network. In addition, $P(k^{(2)})$ denotes the degree distribution $P(k)$ from an original network because $k^{(2)}=k$.

In order to establish a theoretical prediction on the degree distribution of $n$-clique networks, we propose an approximation method based on the statistical method in \cite{Vazquez2005}. We assume that the clustering spectrum $C(k)$ corresponds to the probability that two neighbors of a node with degree $k$ ($\geq 2$) are linked. First, we consider the probability $\phi_n(k)$ that an edge on a node with degree $k$ is eliminated due to the extraction of $n$-clique networks from an original network. For simplicity, we assume that the probability of an edge to be eliminated from a node is independent from the probability of another edge to be eliminated from the same node and the probability of the same edge to be eliminated from a neighbor.
This assumption is a suitable approximation in the case of random graphs \cite{RandGraph} because the probability that there is an edge between two nodes is constant. We show that the approximation is also suitable in the case of arbitrary large-scale graphs (networks) for large $k$ with numerical simulations. Here, we focus on a subset which consists a node with degree $k$, neighboring nodes and edges among these nodes. Then, the edge on the node with degree $k$ belongs to ${k-1 \choose n-2}$ $n$-cliques which are formed with the probability $C(k)^{n_p}$, where $n_p=(n-1)(n-2)/2$.
That is, the probability that the edge is not contained in one of ${k-1 \choose n-2}$ $n$-cliques is $\left[1-C(k)^{n_p}\right]$. Since the edge is eliminated if the edge is contained in no $n$-cliques, from the assimptation of independence, the probability $\phi_n(k)$ can be written as
\begin{equation}
\phi_n(k)=\left\{1-C(k)^{n_p}\right\}^{{k-1 \choose n-2}}.
\label{eq:phi(k)}
\end{equation} 

Next, we characterize the conditional probability that the degree shifts from $k$ to $k^{(n)}$ due to the extraction of $n$-clique networks using the probability $\phi_n(k)$. The conditional probability can be expressed using the bimodal formula, and we have
\begin{equation}
\Phi_n(k^{(n)}|k)={k \choose k^{(n)}}[1-\phi_n(k)]^{k^{(n)}}\phi_n(k)^{(k-k^{(n)})}.
\label{eq:Phi(k|j)}
\end{equation}

The degree distribution from an $n$-clique network $P(k^{(n)})$ is proportional to the sum of $P(k)\Phi_n(k^{(n)}|k)$ for $k=k^{(n)},\ k^{(n)}+1,\ \cdots\ ,\ k_{max}$. Therefore, the degree distribution is finally described as
\begin{equation}
P(k^{(n)})=\frac{N}{N_n}\sum_{k=k^{(n)}}^{k_{max}}P(k)\Phi_n(k^{(n)}|k),
\label{eq:P_n(k)}
\end{equation}
where $N$ and $N_n$ correspond to the total number of nodes in an original network and in a $n$-clique network, respectively. Using $P(k)$ and $C(k)$, the total number of nodes in the $n$-clique network can be estimated by
\begin{equation}
N_n=N\sum_{k=n-1}^{k_{max}}P(k)\left[1-\left\{1-C(k)^{n_p}\right\}^{{k \choose n-1}}\right].
\label{eq:N_n}
\end{equation} 

In order to confirm the theoretical predictions, we performed numerical simulations for the Barab\'asi-Albert (BA) network \cite{Barabasi1999-2}, which provides power-law degree distribution; $P(k)\sim k^{-\gamma}$ with the degree exponent $\gamma=3$.
Figure \ref{fig:deg_BA} shows the degree distributions of $n$-clique networks embedded in the BA network. As shown in Fig. \ref{fig:deg_BA}, our theoretical predictions are in good agreement with the numerical results, indicating that the approximation is suitable.
In addition, the different degree distributions are observed between the $n$-clique networks and the original network.

\section{Shift of the degree}
The degree at a node shifts due to the extraction $n$-clique networks from an original network. Here, we consider the theoretical predictions for the shifts with the statistical properties from an original network. Using the probability $\phi_n(k)$ [Eq. (\ref{eq:phi(k)})] that an edge is eliminated due to the extraction of $n$-clique networks, the expectation value of the degree at a node in a $n$-clique network can be written as
\begin{equation}
k^{(n)}=k\left[1-\phi_n(k)\right].
\label{eq:k'}
\end{equation} 
The probability $\phi_n(k)$ is dependent on the clustering spectrum $C(k)$ as shown in Eq. (\ref{eq:phi(k)}). Since it is empirically found that the spectrum follows the power law in most complex networks \cite{Ravasz2003}, we assume the power-law spectrum; hence $C(k)=C_0k^{-\alpha}$. Moreover, we use the feature of Napier's number, $e^{-c}=(1-c/k)^k$ for large $k$, to rewrite the probability $\phi_n(k)$ [Eq. (\ref{eq:phi(k)})]. In doing such we have
\begin{equation}
\phi_n(k)=\exp\left[-\frac{C_0^{n_p}}{(n-2)!}k^{\zeta_n}\right],
\label{eq:phi(k)_2}
\end{equation}
where
\begin{equation}
\zeta_n=n-n_p\alpha-2.
\label{eq:D_n}
\end{equation}
In particular, the probability $\phi_n(k)$ is independent of the degree $k$ when $\zeta_n=0$, and the proportional relationship between $k^{(n)}$ and $k$ is satisfied.

\begin{figure}[tbp]
\begin{center}
	\includegraphics{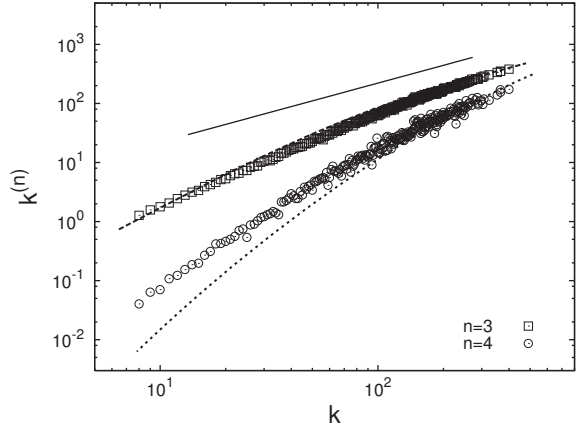}  
	\caption{Shift of the degree at a node due to the extraction of $n$-clique networks from the BA network with $N=3,000$ and $\langle k \rangle=16$. The symbols correspond to the numerical results, and the dashed lines are given by Eq. (\ref{eq:k'}). The solid lines show $k^{(n)}\propto k$.}
	\label{fig:chg_deg}
\end{center}
\end{figure}

\begin{figure}[tbp]
\begin{center}
	\includegraphics{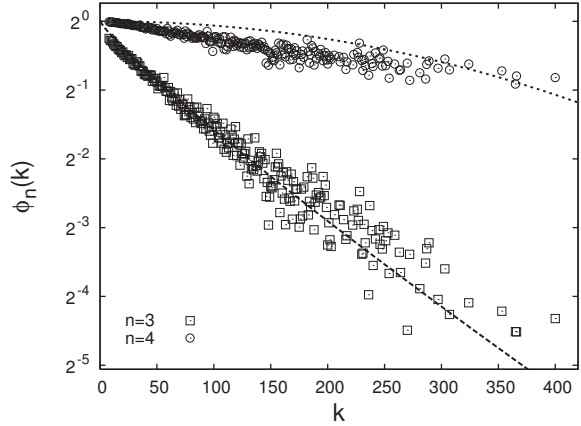}  
	\caption{Probability $\phi_n(k)$ is obtained from the extraction of $n$-clique networks from the BA network with $N=3,000$ and $\langle k \rangle=16$. The symbols correspond to the numerical results, and the dashed lines are given by Eq. (\ref{eq:phi(k)_2}). }
	\label{fig:phi_k}
\end{center}
\end{figure}

In order to confirm the theoretical prediction, we performed numerical simulations for the BA network. Figure \ref{fig:chg_deg} shows the shift of the degree at a node due to the extraction of the $n$-clique networks. As shown in Fig. \ref{fig:chg_deg}, our theoretical prediction is in good agreement with the numerical results. Figure \ref{fig:phi_k} shows the probability $\phi_n(k)$ which is obtained from the extraction of $n$-clique networks. Assume that $C(k)=C_0k^{-\alpha}$, $C_0$ and $\alpha$ are about 0.02 and 0.1 with least-square method, respectively. We give the theoretical prediction with these values. As shown in Eq. (\ref{eq:phi(k)_2}), $\phi_k$ declines exponentially with $k$, indicating that a degree of a high-degree node tends to stay. The prediction is in agreement with the numerical results.

In the case of $n=4$, however, the agreements are weak in Fig. \ref{fig:chg_deg} and \ref{fig:phi_k}. There are two reasons. One is the assumption of independence. In scale-free network, low-degree nodes tend to connect to high-degree nodes. As shown in Fig. \ref{fig:phi_k}, the probability that an edge on the high-degree node is eliminated is very small. For this reason, real $\phi_n(k)$ for small $k$ tends to be smaller than Eq. (\ref{eq:phi(k)_2}). Therefore, real $k^{(n)}$ tends to be larger than our theoretical prediction. Another is fluctuation in clustering spectra $C(k)$. In the case of scale-free networks, the fluctuation is large for small $k$, and is contrary small for large $k$ because of heterogeneous connectivity. And, the probability that a $n$-clique is formed is described as $C(k)^{n_p}$. That is, the error increases with $n_p$. Therefore, our theoretical prediction tends to be in weak agreement in the case of large $n$ and small $k$.

The clustering spectrum of the BA network is independent of the degree $k$ \cite{Barrat2005}. That is, $\alpha\approx 0$. According to Eq. (\ref{eq:D_n}), we predict that the shifts of the degree follow the nonlinear relationship because of the nonzero $\zeta_n$; for example, $\zeta_3=3-2=1$ and $\zeta_4=4-2=2$. As shown in Fig. \ref{fig:chg_deg}, our prediction is in agreement with the numerical results.

\begin{table}[tbp]
\caption{Network sizes, average degrees, and characteristic exponents of the investigated real-world networks and the BA network. The exponents $\gamma$ and $\alpha$ are extracted using the maximum likelihood estimation \cite{Newman2005} and the analytical approximation \cite{Vazquez2004}; thus $C(k)=C_0/\{1+(k/k_0)^\alpha\}$, respectively.}
\label{table:real_net}
\begin{center}
\begin{tabular}{lcccccc}
\hline
\hline
Network & $N$ & $\langle k \rangle$ & $\gamma$ & $\alpha$ & Ref. \\
\hline
Internet (AS level) & 7,832 & 4.38 & 2.4 & 0.75 & \cite{CAIDA} \\
Metabolic ({\it E. coli})& 1,273 & 2.15 & 3.0 & 1.0 & \cite{KEGG} \\
Protein interaction (Yeast)&1,485& 2.62 &2.2 & 1.3 & \cite{Jeong2001} \\
Barab\'asi-Albert & 3,000 & 16.0 & 3.0 & 0.0 & \cite{Barabasi1999-2} \\
\hline
\hline
\end{tabular}
\end{center}
\end{table}

\begin{figure}[tbp]
\begin{center}
	\includegraphics{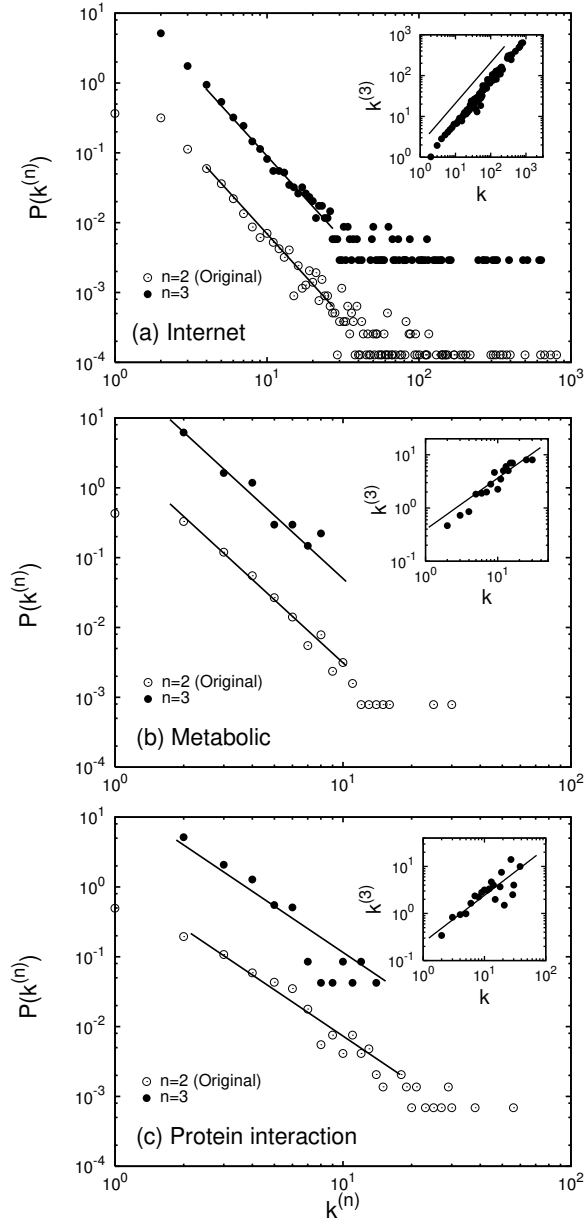}  
	\caption{Degree distributions of $n$-clique networks embedded in the investigated networks (shifted for clarity). The solid lines show $\propto k^{-\gamma}$ in the each main panel. The exponents $\gamma$ are provided from Table \ref{table:real_net}, respectively. The each inset shows the shift of the degree due to the extraction of $3$-clique network. In the each inset, the solid lines correspond to $\propto k$. (a) Internet (AS level), (b) Metabolic network of {\it E. coli}, and (c) protein-protein interaction network of yeast.}
	\label{fig:real_deg}
\end{center}
\end{figure}

\section{Invariance of statistical property}
We discuss statistical properties of $n$-clique networks embedded in a network with power-law statistical properties. Here, we focus on the scale-free connectivity which is one of the well-known power-law statistical properties and is defined as a power-law degree distribution, $P(k)\sim k^{-\gamma}$. 
In networks with scale-free connectivity, we predict that forms of the degree distributions are invariant between the 3-clique and the original network when $\zeta_n=0$. This is because the proportional relationship between the degrees at nodes in the original network and in the $n$-clique networks, is satisfied under this condition.  

In order to verify our prediction, we investigate the degree distributions from $n$-clique networks embedded in several real-world networks with scale-free connectivity: the autonomous system representation of the Internet \cite{CAIDA}, the metabolic network of {\it Escherichia coli} \cite{KEGG}, and the protein-protein interaction network of yeast \cite{Jeong2001}.
These real-world networks have hierarchical modularity, indicating the power-law clustering spectra; hence, $C(k)\sim k^{-\alpha}$ with $\alpha\approx 1$ \cite{Ravasz2003}. In addition, we also consider the BA network, which does not have hierarchical modularity, for comparison. We summarize the networks size, the average degrees, and the exponents characterizing each network in Table \ref{table:real_net}.

The exponents $\alpha$ from the real-world networks with hierarchical modularity are almost one \cite{Ravasz2003} (see also Table \ref{table:real_net}). Therefore, we expect that the forms of the degree distributions are invariant between the 3-clique and the original network because $\zeta_3\approx 3-1-2=0$.
Figure \ref{fig:real_deg} shows the degree distributions of $n$-clique networks embedded in the real-world networks. As expected, the forms of the degree distributions are invariant between the original and the 3-clique network because of the proportional relationship between $k$ and $k^{(3)}$ (see the insets in Fig. \ref{fig:real_deg}).

In contrast, the exponent $\alpha$ from the BA network is equivalent to zero \cite{Barrat2005} (see also Table \ref{table:real_net}) because of there is no hierarchical modularity. Therefore we predict that the power-law degree distribution from an original network is variant due to the extraction of the $3$-clique network (because $\zeta_3=3-2=1$).
Figure \ref{fig:deg_BA} shows the degree distributions of $n$-clique networks embedded in BA networks. As expected, the form of the degree distribution is variant between the 3-clique network and the original network because of the nonlinear relationship between $k$ and $k^{(3)}$ (Fig. \ref{fig:chg_deg}).
 
\section{Discussion and conclusion}
In this paper, we have provided theoretical predictions using the approximation method for the degree distribution of a $n$-clique network and the shifts of the degree due to the extraction of the $n$-clique network. Moreover, we performed numerical simulations and show that the numerical results are in good agreement with our theoretical predictions, indicating that the approximation method is suitable.

Furthermore, we have found that the power-law degree distributions are identical between the 3-clique and the original networks in the scale-free networks with hierarchical modularity using our theoretical predictions. We have only focused on the power-law degree distributions in this paper.
However, because of the proportional relationship between $k$ and $k^{(3)}$, the converse holds for the other power-law statistical properties which are observed in real-world networks: the hierarchical modularity \cite{Ravasz2003} and the assortativity \cite{Newman2002}.

We have confirmed that the power-law statistical properties are invariant between the 3-clique networks and the original networks, although there is no space for the showing of the data. The invariance of the statistical properties implies that structural properties are identical between $3$-clique and original networks. In addition, from these results, we expect that the $3$-clique networks are constructed by the same mechanisms as the original networks with hierarchical modularity.

In contrast, we have found that the 3-clique network embedded in the BA network which does not have hierarchical modularity has different statistical properties from the original network. That is, the structural properties are different between $3$-clique and original networks in the BA network.

We believe that these results provide new insights into global structures of combined network motifs, community structures \cite{Palla2005,Clauset2004} in social and biological networks.
In this paper, expressly, we found structural properties are identical between 3-clique networks and original networks. This lets us expect that 3-clique networks are constructed by the same design principles as the original networks with hierarchical modularity, and it implies that the clique networks help to understand design principles and global structures of combined significant subgraphs which reflect community and functional modules in networks.

For example, it is believed that most real-world networks are constructed by the preferential attachment \cite{Barabasi1999,Reka2002}.
Because of a structural identity between 3-clique networks and original networks, we expect that the clique networks are also constructed by the same preferential attachment as the original networks. This mechanism suggests the preferential attachment of cliques \cite{Takemoto2005}.
Actually, it is reported that there is a preferential attachment of community in social networks \cite{Pollner2006}.
In biological networks, furthermore, cliques correspond to functional modules such as network motifs.
In particular, 3-node clique, which denotes the network motifs such as the feedforward loop and so on, appears frequently.
From our result, we expect that a network which consists of network motifs only is constructed by the same preferential attachment as an original network.
If so, the motifs may concentrate on hubs.
Actually, the concentration of motifs has been found by the network analysis \cite{Vazquez2004}.

In this manner, we believe that we can find new structural properties and new insights into design principles of networks via an analysis of clique networks.
And, our theoretical predictions may help the analysis and its interpretation.
In biological networks, especially, since it is difficult to discuss network formation processes because of no ancestral networks, we believe that the analysis help to understand design principles of networks.
In addition, we may establish more realistic growing network models via the analysis.

\section*{Acknowledgment}
This work was partially supported by Grant-iaZn-Aid No.18740237 from MEXT (JAPAN).

 


\begin{thebibliography}{}

\bibitem{Milo2002} R. Milo {\it et al.}, Science 298 (2002), 824.

\bibitem{Alon2002} S. Shen-Orr, R. Milo, S. Mangan, and U. Alon, Nat. Genet. 31 (2002), 64.

\bibitem{Scotts2000} J. Scotts, Social Network Analysis: A Handbook 2nd edn, Sage, London, 2000. 

\bibitem{Watts2002} D. J. Watts, P. S. Dodds, and M. E. J. Newman, Science 296 (2002), 1302.

\bibitem{Palla2005} G. Palla, I. Der\'enyi, I. Farkas, and T. Vicsek, Nature 435 (2005), 814.

\bibitem{Bianconi2003}G. Bianconi and A. Capocci, Phys. Rev. Lett. 90 (2003), 078701.

\bibitem{Itzkovitz2003} S. Itzkovitz {\it et al.} Phys. Rev. E 68 (2003), 026127.

\bibitem{Vazquez2004} A. V\'azquez {\it et al.}, Proc. Natl. Acad. Sci. U.S.A. 101 (2004), 17940.

\bibitem{Vazquez2005} A. V\'azquez, J. G. Oliveira, and A.-L. Barab\'asi, Phys. Rev. E 71 (2005), 025103(R). 

\bibitem{Costa2004} L. da F. Costa, Phys. Rev. E 70 (2004), 056106.

\bibitem{Derenyi2005} I. Der\'enyi, G. Palla, and T. Vicsek, Phys. Rev. Lett. 94 (2005), 160202.

\bibitem{Barabasi1999} A.-L. Barab\'asi and R. Albert, Science 286 (1999), 509.

\bibitem{Reka2002} R. Albert and A.-L. Barab\'asi, Rev. Mod. Phys. 74 (2002), 47.

\bibitem{Ravasz2003} E. Ravasz and A.-L. Barab\'asi, Phys. Rev. E 67 (2003), 026112.

\bibitem{Takemoto2005} K. Takemoto and C. Oosawa, Phys. Rev. E 72 (2005), 046116.

\bibitem{Skiena1997} S. S. Skiena, The Algorithm Design Manual, Springer-Verlag, New York, 1997.

\bibitem{RandGraph} B. Bollob\'as, Random Graphs, Achademic Press, New York, 1985.

\bibitem{Kashtan2004} N. Kashtan, S. Itzkovitz, R. Milo, and U. Alon, Phys. Rev. E 70 (2004), 031909.

\bibitem{Ishihara2005} S. Ishihara, K. Fujimoto, and T. Shibata, Genes to Cells 10 (2005), 1025.

\bibitem{Barabasi1999-2} A.-L. Barab\'asi, R. Albert, and H. Jeong, Physica A 272 (1999), 173.

\bibitem{Barrat2005} A. Barrat and R. Pastor-Satorras, Phys. Rev. E 71 (2005), 36127.

\bibitem{Newman2005} M. E. J. Newman, Contemp. Phys. 46 (2005), 323.

\bibitem{CAIDA} The Cooperative Association for Internet Data Analysis, located at the San Diego Supercomputer Center, provided macroscopic topology AS adjacencies (see http://www.caida.org/home/).

\bibitem{KEGG} M. Kanehisa {\it et al.}, Nucl. Acids Res. 34 (2003), D354.

\bibitem{Jeong2001} H. Jeong, S. Mason, A.-L. Barab\'asi, and Z. N. Oltvai, Nature 411 (2001), 41.

\bibitem{Newman2002} M. E. J. Newman, Phys. Rev. Lett. 89 (2002), 208701.

\bibitem{Clauset2004} A. Clauset, M. E. J. Newman, and C. Moore, Phys. Rev. E 70 (2004), 066111.

\bibitem{Pollner2006} P. Pollner, G. Palla, and T. Vicsek, Europhys. Lett. 73 (2006), 478. 

\end{thebibliography}
\end{document}